\documentclass[aps,prc,amsmath,preprint,superscriptaddress,showpacs]{revtex4-1}

\usepackage{amsmath}
\usepackage{bm}
\usepackage{graphicx}

\begin{document}


\title{Complete set of polarization transfer observables
for the ${}^{16}{\rm O}(\vec{p},\vec{n}){}^{16}{\rm F}$ 
reaction at 296 MeV and 0 degrees}


\author{T.~Wakasa}
\email[]{wakasa@phys.kyushu-u.ac.jp}
\homepage[]{http://ne.phys.kyushu-u.ac.jp/~wakasa}
\affiliation{Department of Physics, Kyushu University,
Higashi, Fukuoka 812-8581, Japan}
\author{M.~Okamoto}
\affiliation{Department of Physics, Kyushu University,
Higashi, Fukuoka 812-8581, Japan}
\author{M.~Takaki}
\affiliation{Center for Nuclear Study, The University of Tokyo, 
Bunkyo, Tokyo 113-0033, Japan}
\author{M.~Dozono}
\affiliation{Center for Nuclear Study, The University of Tokyo, 
Bunkyo, Tokyo 113-0033, Japan}
\author{K.~Hatanaka}
\affiliation{Research Center for Nuclear Physics, Osaka University,
Ibaraki, Osaka 567-0047, Japan}
\author{M.~Ichimura}
\affiliation{RIKEN Nishina Center for Accelerator-Based Science,
The Institute of Physical and Chemical Research,
Wako, Saitama 351-0198, Japan}
\author{T.~Noro}
\affiliation{Department of Physics, Kyushu University,
Higashi, Fukuoka 812-8581, Japan}
\author{H.~Okamura}
\thanks{Deceased.}
\affiliation{Research Center for Nuclear Physics, Osaka University,
Ibaraki, Osaka 567-0047, Japan}
\author{Y.~Sakemi}
\affiliation{Cyclotron and Radioisotope Center, Tohoku University,
Sendai, Miyagi 980-8578, Japan}


\date{\today}

\begin{abstract}
 We report measurements of the cross section and 
a complete set of polarization transfer observables 
for the ${}^{16}{\rm O}(\vec{p},\vec{n}){}^{16}{\rm F}$ reaction 
at a bombarding energy of $T_p$ = 296 MeV and 
a reaction angle of $\theta_{\rm lab}$ = $0^{\circ}$.
 The data are compared with distorted-wave 
impulse approximation calculations employing 
the large configuration-space shell-model (SM) wave functions.
 The well-known Gamow-Teller and spin-dipole (SD) states 
at excitation energies of $E_x$ $\lesssim$ 8 MeV 
have been reasonably reproduced by the calculations
except for the spin--parity $J^{\pi}$ = $2^-$ 
state at $E_x$ = 5.86 MeV.
 The SD resonance at $E_x$ $\simeq$ 9.5 MeV appears to 
have more $J^{\pi}$ = $2^-$ strength than $J^{\pi}$ = $1^-$ 
strength, consistent with the calculations.
 The data show significant strength in the spin-longitudinal 
polarized cross section $ID_L(0^{\circ})$ at $E_x$ $\simeq$ 15 MeV, 
which indicates existence of the $J^{\pi}$ = $0^-$ SD resonance as 
predicted in the SM calculations.
\end{abstract}


\maketitle

\section{INTRODUCTION}
\label{sec:intro}

 The details of spin excitations in nuclei remain interesting and stimulating 
problems in a variety of aspects \cite{giant_resonances}.
 In particular, quenching of the Gamow-Teller (GT) 
($L=0$, $S=1$, $J^{\pi}$ = $1^+$)
strength in nuclei has been 
the subject of intensive theoretical and experimental investigation 
\cite{ppnp_56_446_2006}.
 The $(p,n)$ and $(n,p)$ reactions in the intermediate energy 
region have been found to be extremely useful probes for
studying the spin--isospin $\bm{\sigma}\bm{\tau}$ 
correlations in nuclei with refined accuracy.
 Recent experimental studies \cite{prc_55_2909_1997,plb_615_193_2005}
have revealed that 
GT quenching is mainly caused by coupling to 
two-particle--two-hole ($2p$--$2h$) excitations, 
while the $\Delta$--hole coupling 
plays a minor role.

 Spin-dipole (SD) 
($L=1$, $S=1$, $J^{\pi}$ = $0^-$, $1^-$, and $2^-$)
excitations have also been studied extensively in experimental 
studies.
 Theoretical investigations of SD excitations give rise to
interesting problems, especially in relation to
nuclear structure \cite{prc_30_1032_1984,npa_435_1_1985,prc_57_139_1998}
and astrophysical considerations \cite{prl_76_2629_1996}.
 In the $A$ = 12 system, 
the SD resonances (SDRs) were found to occur at excitation energies of  
$E_x$ $\simeq$ 4 and 7 MeV.
 The former SDR is assigned as $J^{\pi}=2^-$ \cite{npa_506_1_1990}
while the latter SDR is considered to be mainly 
$J^{\pi}=1^-$ from the studies of the cross sections 
for the 
${}^{12}{\rm C}(p,n){}^{12}{\rm N}$ 
\cite{prc_52_2535_1995,prc_54_237_1996} and 
${}^{12}{\rm C}(n,p){}^{12}{\rm B}$ 
\cite{npa_559_368_1993,prc_48_1158_1993} reactions.
 The $J^{\pi}=1^-$ dominance for the SDR at $E_x$ $\simeq$ 7 MeV 
has been supported by measurements of 
both the proton decay
of the SDR in ${}^{12}{\rm N}$ populated by 
the ${}^{12}{\rm C}({}^{3}{\rm He},t){}^{12}{\rm N}$
reaction at ${}^{3}{\rm He}$ incident energies of 
$T_{{}^{3}{\rm He}}$ = 75 and 81 MeV \cite{npa_405_109_1983}
and the neutron decay 
of the SDR in ${}^{12}{\rm B}$ populated by
the ${}^{12}{\rm C}(d,{}^{2}{\rm He}){}^{12}{\rm B}$
reaction at a deuteron incident energy of 
$T_d$ = 200 MeV \cite{prc_57_3153_1998}.
 However, measurement of the tensor analyzing 
power for the ${}^{12}{\rm C}(\vec{d},{}^{2}{\rm He}){}^{12}{\rm B}$
reaction at $T_d$ = 270 MeV 
suggests that the SDR 
mainly consists of $J^{\pi}=2^-$ \cite{plb_345_1_1995}. 
 This result has been supported by 
measurement of the complete set of polarization observables 
for the ${}^{12}{\rm C}(\vec{p},\vec{n}){}^{12}{\rm N}$ 
reaction at a proton incident energy of $T_p$ = 296 MeV and 
a reaction angle of $\theta_{\rm lab}$ = $0^{\circ}$ 
\cite{jpsj_77_014201_2008}.
 The later high-resolution measurement for
the ${}^{12}{\rm C}(\vec{d},{}^{2}{\rm He}){}^{12}{\rm B}$ 
reaction at $T_d$ = 171 MeV \cite{plb_649_35_2007} reveals 
that the low- and high-energy parts of the SDR at $E_x$ $\simeq$ 7 MeV 
mainly consists 
of $J^{\pi}=2^-$ and $1^-$ strengths, respectively.
 Similar conclusions are made by Inomata {\it et al.} 
\cite{prc_57_3153_1998}
from measurement of the proton decay of the SDR in 
${}^{12}{\rm N}$ produced by 
the ${}^{12}{\rm C}({}^{3}{\rm He},t){}^{12}{\rm N}$
reaction at 
$T_{{}^{3}{\rm He}}$ = 450 MeV.
 They also conclude that the high-energy part of the SDR at 
$E_x$ $\simeq$ 4 MeV in ${}^{12}{\rm N}$ mainly consists 
of $J^{\pi}=1^-$ by the same measurement, 
suggesting the fragmentation of the $J^{\pi}=1^-$ strength.
 This fragmentation has been supported by
theoretical calculations including the tensor correlation 
\cite{npa_637_547_1998} and the deformation effect 
\cite{sgr97_kurasawa}.

 Another long-standing problem in relation to SD excitations 
in such systems
is the missing $J^{\pi}=0^-$ strength.
 In the $A=12$ system, shell-model (SM) calculations 
predict a fairly large 
$J^{\pi}$ = $0^-$ SD state at $E_x$ $\simeq$ 8--9 MeV.
 Extensive experimental efforts have been made to 
identify this $J^{\pi}$ = $0^-$ state by measuring the cross section,
however, clear evidence was not obtained.
 Recently, the tensor analyzing powers 
in the $(\vec{d},{}^{2}{\rm He})$ reaction 
\cite{prc_66_054602_2002,plb_649_35_2007} 
and the polarization 
transfer observables in the $(\vec{p},\vec{n})$ reaction
\cite{jpsj_77_014201_2008} have been measured.
 The results of these measurements suggest the existence 
of $J^{\pi}$ = $0^-$ states at $E_x$ = 9.3 and 8.4 MeV 
in ${}^{12}{\rm B}$ and ${}^{12}{\rm N}$, respectively.

 For the $A$ =16 system, the $J^{\pi}$ = $0^-$ and $1^-$ SD strengths 
were also found to be missing in a study of the tensor analyzing 
powers for the ${}^{16}{\rm O}(\vec{d},{}^{2}{\rm He}){}^{16}{\rm N}$ 
reaction at $T_{d}$ = 270 MeV \cite{spin2004_suda}.
 Evidence for the missing $0^-$ state predicted by the 
SM calculations was suggested in a study of the 
${}^{16}{\rm O}(\vec{p},\vec{n}){}^{16}{\rm F}$ reaction 
at $T_p$ = 135 MeV \cite{npa_577_79c_1994}, 
however, this has not been settled.
 It should be noted that the SD excitations in ${}^{16}{\rm O}$ 
have been discussed in relation to neutrino detection 
from supernovae at the Super-Kamiokande water Cherenkov 
detector \cite{prl_76_2629_1996}.
 Thus it is very important to obtain quantitative information 
on the distribution of the SD strengths in such systems.

 In this article, we present the double-differential 
cross section and a complete set of polarization transfer 
observables for the ${}^{16}{\rm O}(\vec{p},\vec{n}){}^{16}{\rm F}$ 
reaction at $T_p$ = 296 MeV and $\theta_{\rm lab}$ = $0^{\circ}$.
 It should be noted that the SD states are fairly strongly 
excited even at $\theta_{\rm lab}$ = $0^{\circ}$ 
\cite{npa_577_79c_1994} because the 
GT transitions are largely inhibited for a 
double $LS$ closed shell nucleus as occurs in ${}^{16}{\rm O}$.
 In addition, distortion effects are minimal at  $T_p$ $\simeq$ 300 MeV
\cite{ppnp_56_446_2006}, thereby enabling the extraction of reliable 
nuclear structure information on the SDRs.
 Polarization transfer observables
are sensitive to the spin--parity 
of an excited state \cite{prc_26_727_1982}, 
as was demonstrated for the SDR in 
${}^{12}{\rm N}$ \cite{jpsj_77_014201_2008}.
 They are used to separate the cross section into 
non-spin, $ID_0(0^{\circ})$, 
spin-longitudinal, $ID_L(0^{\circ})$, and 
spin-transverse, $ID_T(0^{\circ})$, polarized cross sections. 
 The observed $ID_i(0^{\circ})$ are compared 
with distorted-wave 
impulse approximation (DWIA) calculations employing the 
large configuration-space SM wave functions 
\cite{prc_65_024322_2002} 
in order to access the spin--isospin excitations in 
${}^{16}{\rm F}$, e.g., 
the missing $J^{\pi}$ = $0^-$ and $1^-$ SD strengths.

\section{EXPERIMENTAL METHODS}
\label{sec:exp}

 Measurements were performed with 
a neutron time-of-flight (NTOF) system \cite{nima_369_120_1996} 
and 
a neutron detector and polarimeter (NPOL3) \cite{nima_547_569_2005} 
at the Research Center for Nuclear Physics (RCNP) at Osaka University.
 Detailed descriptions of the NTOF and NPOL3 systems 
are found in 
Refs.~\cite{nima_369_120_1996,nima_547_569_2005}.
 Thus, in the following, we only describe
the detector system briefly and discuss experimental details relevant
to the present experiments.

\subsection{Polarized proton beam}

 The polarized proton beam from the 
high-intensity polarized ion source (HIPIS) at RCNP 
\cite{nima_384_575_1997} was accelerated up to 
$T_p$ = 53 and 296 MeV by using the AVF and 
Ring cyclotrons, respectively.
 The beam polarization direction was reversed every 5 s 
by selecting rf transitions in order to minimize 
geometrical false asymmetries.
 For the cross section measurements, 
one out of seven beam pulses was selected for injection into 
the Ring cyclotron, 
which then yielded a beam pulse period of 453 ns.
 This pulse selection reduces the wraparound of slow 
neutrons from preceding beam pulses.
 For the polarization transfer measurements, pulse 
selection was not performed in order to achieve 
reasonable statistical accuracy.
 In both measurements, single-turn extraction from 
the Ring cyclotron was used
in order to maintain the beam polarization.

 The superconducting solenoid magnets, SOL1 and SOL2 
\cite{nima_369_120_1996},
were located in the injection line from the AVF to the Ring 
cyclotrons to precess the proton spin direction.
 Each magnet can rotate the direction of the polarization 
vector from the normal $\hat{N}$ into 
the sideways $\hat{S}$ directions.
 The two magnets were installed in front of (SOL1) and behind 
(SOL2) the $45^{\circ}$ bending magnet, 
and the spin precession angle in this 
bending magnet was about $85.2^{\circ}$ for $T_p$ = 53 MeV 
protons.
 Therefore, we can obtain proton beams 
with longitudinal ($\hat{L}$) and sideways ($\hat{S}$) 
polarizations at the exit of the SOL2 
by using the SOL1 and SOL2 magnets, respectively.

 The beam polarization was continuously monitored
by two sets of beam-line polarimeters, BLP1 and BLP2 
\cite{nima_369_120_1996}, which were installed 
in front of and behind the $98^{\circ}$ bending magnet, 
respectively.
 Each polarimeter consists of four conjugate-angle pairs of 
plastic scintillators, and determines the beam polarization 
via the $\vec{p}+p$ elastic scattering in the 
$\hat{N}$ and $\hat{S}$ directions.
 A self-supporting polyethylene (${\rm CH_2}$) target with a 
thickness of 1.1 ${\rm mg/cm^2}$ was used as the hydrogen target, 
and the elastically scattered and recoiled protons were detected 
in kinematical coincidence with a pair of scintillators.
 The spin precession angle in the $98^{\circ}$ bending magnet 
was about $231.1^{\circ}$ for $T_p$ = 296 MeV protons.
 Therefore, all components $(p_S, p_N, p_L)$  
of the polarization vector can be simultaneously determined 
using BLP1 and BLP2.
 The typical magnitude of the beam polarization was about 0.62.

\subsection{${}^{16}{\rm O}$ target}

 The ${}^{16}{\rm O}$ target was prepared as a windowless 
and self-supporting ice (${\rm H_2O}$) target 
\cite{nima_459_171_2001}.
 This target was operated at temperatures down to 77 K 
by using liquid nitrogen, and the typical areal density was 
about 147 ${\rm mg/cm^2}$.
 The thickness was determined by comparing the 
${}^{16}{\rm O}(p,n){}^{16}{\rm F}$ yield to that from a ${\rm SiO_2}$ target 
with a thickness of 221 ${\rm mg/cm^2}$.
 Since the hydrogen does not produce any physical background 
in the present energy region, we have successfully obtained 
very clean spectra for the ${}^{16}{\rm O}(p,n){}^{16}{\rm F}$ 
reaction.

\subsection{Neutron spin-rotation magnet and NPOL3}

 A dipole magnet (NSR magnet) was positioned at the entrance
of the time-of-flight (TOF) tunnel.
 This magnet was used to precess the neutron polarization 
vector from the longitudinal direction, $\hat{L}'$, 
to the normal direction, $\hat{N}'$, 
so as to allow the longitudinal component 
to be measured with NPOL3 as the normal component.

 Neutrons were measured by the NPOL3 system \cite{nima_547_569_2005} 
with a 100~m flight path length.
 The NPOL3 system consists of 
three planes of neutron detectors.
 Each of the first two planes (HD1 and HD2) 
consists of 10 sets of 
one-dimensional position-sensitive plastic scintillators (BC408) 
with a size of 100 $\times$ 10 $\times$ 5 ${\rm cm}^3$.
 Each plane has an effective detection area of 1 ${\rm m^2}$.
 The last plane (NC) is a two-dimensional position-sensitive 
plastic scintillator 
with a size of 100 $\times$ 100 $\times$ 10 ${\rm cm}^3$.
 Both HD1 and HD2 planes serve as neutron detectors and neutron polarization 
analyzers for the cross section and polarization transfer measurements, 
respectively, and NC plane acts as a catcher for the 
particles scattered by the HD1 or HD2 plane.

\section{DATA REDUCTION}
\label{sec:reduction}

\subsection{Neutron detection efficiency}

 The neutron detection efficiency of NPOL3 
(HD1 and HD2) was determined 
using the ${}^{12}{\rm C}(p,n){}^{12}{\rm N}({\rm g.s.},1^+)$ 
reaction at $\theta_{\rm lab}$ = $0^{\circ}$, 
which has a known cross section at $T_p$ = 296 MeV 
\cite{plb_656_38_2007,prc_80_024319_2009}.
 The result was 0.048$\pm$0.003 
with the overall uncertainty mainly coming from 
uncertainties in the cross section 
and thickness of the ${}^{12}{\rm C}$ target.

\subsection{Effective analyzing power}

 The neutron polarization was analyzed by monitoring $\vec{n}+p$
scattering at either neutron detector HD1 or HD2, and the recoiled 
protons were detected with neutron detector NC.
 The effective analyzing power $A_{y;{\rm eff}}$ of NPOL3 
was determined by using polarized neutrons from the 
GT transition in the 
${}^{12}{\rm C}(p,n){}^{12}{\rm N}({\rm g.s.},1^+)$ 
reaction at 
$T_p$ = 296 MeV and $\theta_{\rm lab}$ = $0^{\circ}$.
 We used two kinds of polarized protons 
with normal ($p_N$) and longitudinal ($p_L$)
polarizations.
 The corresponding neutron polarizations at $0^{\circ}$ 
become $p_N'=p_ND_{NN}(0^{\circ})$ and 
$p_L'=p_L D_{LL}(0^{\circ})$, respectively.
 The resulting asymmetries measured by NPOL3 
are
\begin{subequations}
\label{eq_asym}
\begin{eqnarray}
\epsilon_N & = & 
   p_N'A_{y;{\rm eff}}=p_ND_{NN}(0^{\circ})A_{y;{\rm eff}} ,
\label{eq_asym_dnn}\\
\epsilon_L & = & 
   p_L'A_{y;{\rm eff}}=p_LD_{LL}(0^{\circ})A_{y;{\rm eff}} .
\label{eq_asym_dll}
\end{eqnarray}
\end{subequations}
 Because the polarization transfer observables for the GT
transition satisfy \cite{jpsj_73_1611_2004}
\begin{equation}
2D_{NN}(0^{\circ})+D_{LL}(0^{\circ})=-1 ,
\label{eq:gt_dii}
\end{equation}
$A_{y;{\rm eff}}$ can be expressed in terms of
Eqs.~(\ref{eq_asym}) and (\ref{eq:gt_dii}) as 
\begin{equation}
A_{y;{\rm eff}} = -\left(
2\frac{\epsilon_N}{p_N}+\frac{\epsilon_L}{p_L}
\right) .
\end{equation}
 Therefore, the $A_{y;{\rm eff}}$ value can be obtained without
knowing a priori the values of $D_{ii}(0^{\circ})$, giving a
result of $A_{y;{\rm eff}}$ = 0.131 $\pm$ 0.004,
in which the uncertainty is statistical.

 The $D_{NN}(0^{\circ})$ value at $T_p$ = 296 MeV, which is
determined from Eq.~(\ref{eq_asym_dnn}) using the obtained value for
$A_{y;{\rm eff}}$, is 
$D_{NN}(0^{\circ})$ = $-0.216\pm 0.013$.
 This value for $D_{NN}(0^{\circ})$ is consistent with a previous value
of $D_{NN}(0^{\circ})$ = $-0.227\pm 0.010$ \cite{prc_80_024319_2009}, 
demonstrating the reliability of our calibrations.

\section{RESULTS}
\label{sec:results}

\subsection{Cross section and polarization transfer observables}

\begin{figure}
\begin{center}
\includegraphics[width=0.6\linewidth,clip]{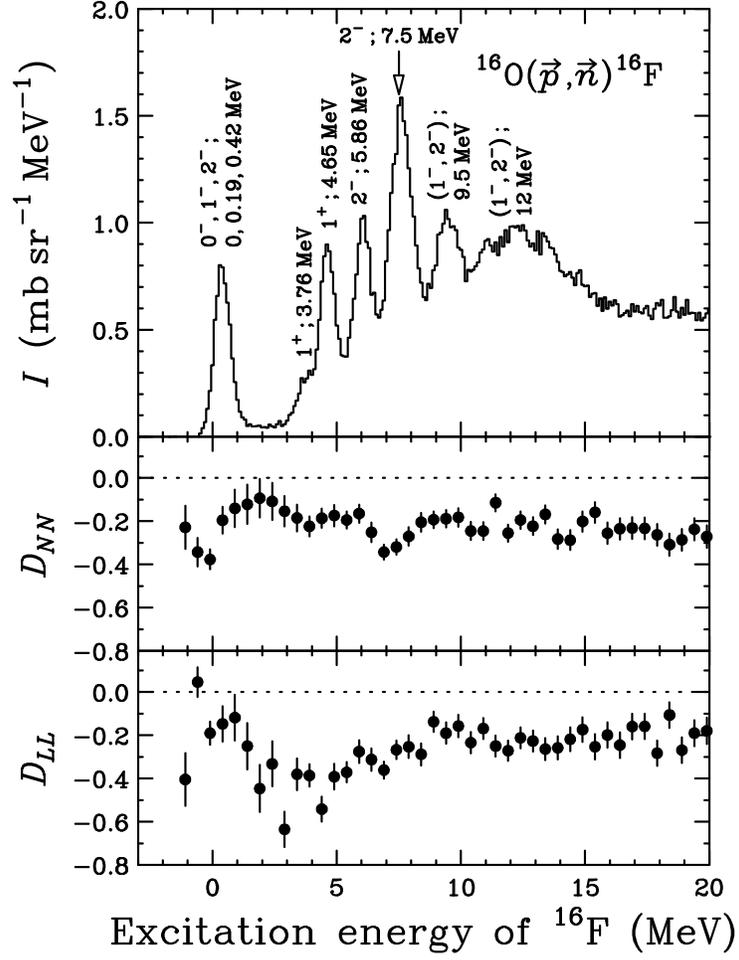}
\end{center}
\caption{
 The double-differential cross section spectrum, $I$ (top panel), and 
a complete set of polarization transfer observables, 
$D_{NN}(0^{\circ})$ (middle panel) and 
$D_{LL}(0^{\circ})$ (bottom panel), 
for the ${}^{16}{\rm O}(\vec{p},\vec{n}){}^{16}{\rm F}$ reaction at 
$T_p$ = 296 MeV and $\theta_{\rm lab}$ = $0^{\circ}$.}
\label{fig1}
\end{figure}

 Figure~\ref{fig1} shows the double-differential cross section $I$ 
and the complete set of polarization 
transfer observables $D_{NN}(0^{\circ})$ and $D_{LL}(0^{\circ})$
for the ${}^{16}{\rm O}(\vec{p},\vec{n}){}^{16}{\rm F}$ reaction 
at $T_p$ = 296 MeV and $\theta_{\rm lab}$ = $0^{\circ}$.
 The data for the cross section are binned in 0.1 MeV intervals, 
while the data for $D_{ii}(0^{\circ})$ are binned in 
0.5 MeV intervals to reduce statistical fluctuations.
 Excitation of the well-known GT and SD states \cite{npa_564_1_1993} 
at $E_x$ $\lesssim$ 8 MeV can be seen.
 The peak at $E_x$ $\simeq$ 0 MeV is a sum of the $J^{\pi}$ = 
$0^-$, $1^-$ and $2^-$ states, while the shoulder at $E_x$ = 3.76 MeV and the peak 
at $E_x$ = 4.65 MeV are the $J^{\pi}$ = $1^+$ states.
 Both the peak at $E_x$ = 5.86 MeV and the narrow resonance at 
$E_x$ $\simeq$ 7.5 MeV are known as $J^{\pi}$ = $2^-$ states.
 The other narrow and broad resonances at $E_x$ $\simeq$ 9.5 and 12 MeV, 
respectively, have been suggested to be features of 
the $J^{\pi}$ = $1^-$ and $2^-$ states \cite{npa_564_1_1993}.

 An interesting feature of the $D_{ii}(0^{\circ})$ data is that negative 
values are obtained over the entire excitation region.
 It should be noted that the $D_{NN}(0^{\circ})$ value of a 
natural-parity transition is predicted to be positive in the 
plane-wave impulse approximation (PWIA) theory \cite{prc_26_727_1982}.
 Thus the observed negative $D_{NN}(0^{\circ})$ values for the 
resonances at $E_x$ $\simeq$ 9.5 and 12 MeV 
indicate significant unnatural-parity 
contributions such as from $J^{\pi}$ = $2^-$ and $1^+$ states, which is consistent with 
a previous result obtained for the 
${}^{16}{\rm O}({}^{3}{\rm He},t){}^{16}{\rm F}$ reaction at 
$T_{{}^{3}{\rm He}}$ = 81 MeV \cite{npa_420_257_1984}.

\subsection{Polarized cross sections}

 The double-differential cross section $I$ can be separated
into non-spin, $ID_0$, spin-longitudinal, $ID_q$, and 
two spin-transverse, $ID_n$ and $ID_p$, polarized cross sections 
as follows:
\begin{equation}
I = ID_0 + ID_q + ID_n + ID_p ,
\end{equation}
where $D_i$ are the polarization observables 
introduced by Bleszynski {\it et al.} \cite{prc_26_2063_1982}.
 Here we also use the spin-longitudinal $ID_L(0^{\circ})$ and 
spin-transverse $ID_T(0^{\circ})$ polarized cross sections, which 
are defined at $\theta_{\rm lab} = 0^{\circ}$ as 
\cite{jpsj_77_014201_2008}
\begin{subequations}
\begin{eqnarray}
ID_L(0^{\circ}) & \equiv & ID_q(0^{\circ}) 
= \frac{I}{4}\left[1-2D_{NN}(0^{\circ})+D_{LL}(0^{\circ})\right] , \\
ID_T(0^{\circ}) & \equiv & ID_n(0^{\circ}) + ID_p(0^{\circ}) 
= \frac{I}{2}\left[1-D_{LL}(0^{\circ})\right] .
\end{eqnarray}
\end{subequations}

\begin{figure}
\begin{center}
\includegraphics[width=0.7\linewidth,clip]{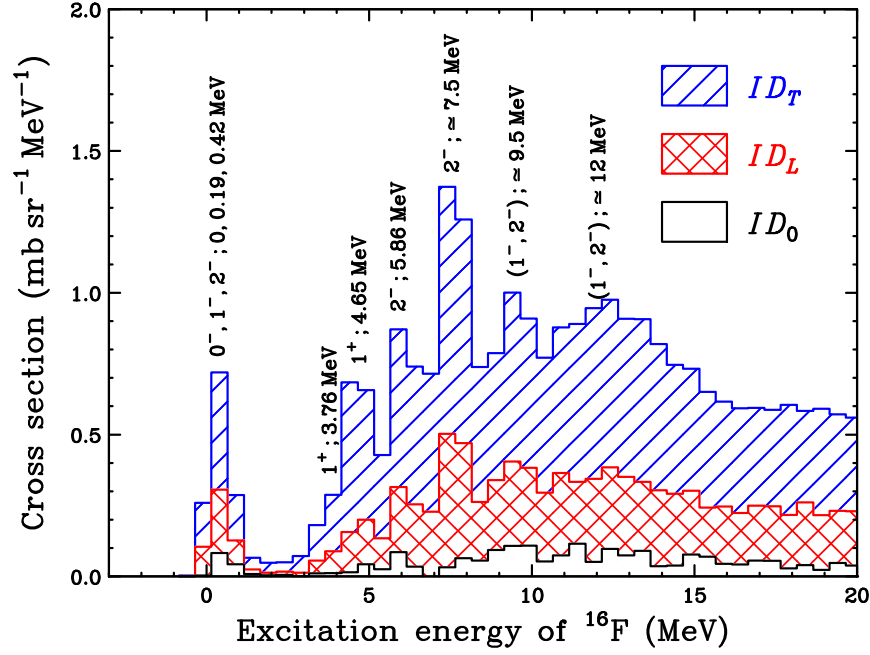}
\end{center}
\caption{(Color online) 
 The non-spin $ID_0$ (unfilled), spin-longitudinal $ID_L$ 
(cross hatched), and spin-transverse $ID_T$ (hatched) 
polarized cross sections 
for the ${}^{16}{\rm O}(\vec{p},\vec{n}){}^{16}{\rm F}$ reaction at 
$T_p$ = 296 MeV and $\theta_{\rm lab}$ = $0^{\circ}$.}
\label{fig2}
\end{figure}

 Figure~\ref{fig2} shows the polarized cross sections, 
$ID_L(0^{\circ})$, $ID_T(0^{\circ})$, and $ID_0(0^{\circ})$, 
for the ${}^{16}{\rm O}(p,n){}^{16}{\rm F}$ reaction 
at $T_p$ = 296 MeV and $\theta_{\rm lab}$ = $0^{\circ}$.
 The data are binned in 0.5 MeV intervals to reduce statistical 
fluctuations.
 The spin-longitudinal cross section, $ID_L(0^{\circ})$, consists 
exclusively of unnatural-parity transitions 
such as $J^{\pi}$ = $1^+$ and $2^-$, 
whereas the spin-transverse cross section, $ID_T(0^{\circ})$, 
consists of both the natural- and unnatural-parity transitions 
\cite{prc_26_727_1982}.
 Note that the unnatural-parity $J^{\pi}$ = $0^-$ 
transition is a special case 
and it contributes to $ID_L(0^{\circ})$ only.
 The peaks and resonances at $E_x$ $\lesssim$ 8 MeV are observed for 
both $ID_L(0^{\circ})$ and $ID_T(0^{\circ})$, which is consistent with 
their unnatural-parity assignments 
to either $J^{\pi}$ = $2^-$ or $1^+$ states.
 At $E_x$ $\simeq$ 9.5 MeV, resonances are observed 
for both $ID_L(0^{\circ})$ and $ID_T(0^{\circ})$ 
while at $E_x$ $\simeq$ 12 MeV only $ID_T(0^{\circ})$ 
displays a resonance.
 These results suggest that the dominant components 
for the resonances at $E_x$ $\simeq$ 9.5 and 
12 MeV are the $J^{\pi}$ = $2^-$ and $1^-$ states, respectively, 
as will be discussed
in greater detail in the next section in relation to the DWIA calculations.

 It is interesting to note that the sum of the $J^{\pi}$ = 
$0^-$, $1^-$, and $2^-$ SD states at $E_x$ $\simeq$ 0 MeV 
forms a significant peak in the $ID_0(0^{\circ})$ spectrum.
 In PWIA theory, it is considered that a SD state 
could not contribute to $ID_0(0^{\circ})$ due to its 
spin-flip character.
 However, the $J^{\pi}$ = $1^-$ state may be apparent 
due to distortion effects \cite{jpsj_77_014201_2008}.
 Therefore, the $J^{\pi}$ = $1^-$ component is considered 
to give rise to the peak 
in the $ID_0(0^{\circ})$ spectrum, 
which will also be investigated in the next section.

\section{\label{sec:method}DISCUSSION}

\subsection{DWIA calculations}

 DWIA calculations were performed on the data using a computer code {\sc dw81} 
\cite{dw70,*npa_97_572_1967,*dw81}.
 The one-body density matrix elements (OBDMEs) for the 
${}^{16}{\rm O}(p,n){}^{16}{\rm F}$ reaction were obtained 
from the SM calculations \cite{prc_65_024322_2002}, 
which were performed 
in the $0s$-$0p$-$1s0d$-$0f1p$ configuration space 
by using phenomenological effective interactions.
 In the calculations, the ground state of ${}^{16}{\rm O}$ 
was described as a mixture of $0\hbar\omega$ (closed-shell) 
and $2\hbar\omega$ configurations, and up to $3\hbar\omega$ 
configurations were included in the final states.
 The $2\hbar\omega$ admixture in the ground state provides 
significant GT strength which is similar to that obtained by 
Haxton and Johnson \cite{prl_65_1325_1990}, 
and the transition strengths for negative-parity states 
are uniformly reduced by a factor of about 0.7 
\cite{prc_65_024322_2002}.
 The single particle wave functions were generated by 
the sum of a Woods-Saxon (WS) potential with 
$r_0$ = 1.27 fm, $a_0$ = 0.67 fm \cite{bohr_mottelson}, 
a spin-orbit potential with $V_{ls}$ = 10.4 MeV 
\cite{prc_51_269_1995}, and the Coulomb potential.
 The depth of the WS potential  was adjusted to reproduce 
the separation energies of the $0p_{1/2}$ orbits.
 The unbound single particle states were assumed to have 
a shallow binding energy of 0.01 MeV in order to simplify the calculations.
 The distorted wave for the protons was generated using a 
global optical model potential (OMP) in the proton energy 
range of $T_p$ = 20--1040 MeV \cite{prc_47_297_1993}, 
while that for neutrons was generated 
using a global OMP in the neutron energy range of 
$T_n$ = 20--1000 MeV \cite{prc_43_2773_1991}.

\subsection{Effective NN interactions}

 The polarization transfer observables $D_{ij}$ are sensitive to 
both the spin-parity of the excited state and the effective 
nucleon--nucleon ({\it NN}\,) interaction.
 Thus, in order to use the 
$D_{ij}$ values for the spin-parity assignments, 
we have checked and modified the effective {\it NN} interaction 
parameterized by Franey and Love at 325 MeV \cite{prc_31_488_1985}. 
 For this purpose, 
the experimental data for well-resolved 
$J^{\pi}$ = $1^+$ (g.s.) and $2^-$ (4.14 MeV) states of the 
${}^{12}{\rm C}(\vec{p},\vec{n}){}^{12}{\rm N}$ reaction at the 
same energy \cite{plb_656_38_2007,prc_80_024319_2009} were compared 
with the DWIA calculations.
 In the calculations, the OBDMEs were obtained by using 
a computer code {\sc Nushell@MSU} \cite{nushell} with the 
PSDMKII interaction \cite{npa_255_315_1975} 
in the $(0+1)\hbar\omega$ configuration space.

 It was found that the $D_{ii}(0^{\circ})$ values for the 
$J^{\pi}$ = $1^+$ state are reasonably reproduced by the 
calculations, which is consistent with previous results 
\cite{plb_656_38_2007,prc_80_024319_2009}.
 However, those for the $J^{\pi}$ = $2^-$ state could not be 
reproduced, e.g., 
the experimental data was 
$D_{LL}(0^{\circ}) = -0.36\pm 0.09$ while the theoretical value was 
$-0.85$.
 It should be noted that the $D_{ii}(0^{\circ})$ values 
are sensitive to the tensor component of the 
interaction \cite{prc_51_r2871_1995,prc_51_3162_1995,plb_459_61_1999}.
 Thus we tried to modify the tensor component 
to reproduce the experimental data.

 The isovector $V_{\tau}^T$ and isoscalar $V_{0}^T$ 
exchange tensor interactions 
are described with the tensor-even $V^{\rm TNE}$ and tensor-odd 
$V^{\rm TNO}$ interactions as 
\cite{prc_24_1073_1981,prc_51_r2871_1995}
\begin{subequations}
\begin{eqnarray}
V_{\tau}^T & = & -\frac{1}{4}(V^{\rm TNE}+V^{\rm TNO}) , \\
V_{0}^T    & = &  \frac{1}{4}(V^{\rm TNE}-3V^{\rm TNO}) .
\end{eqnarray}
\end{subequations}
 We modified the isovector $V_{\tau}^T$ interaction 
while keeping the isoscalar $V_{0}^T$ tensor interaction 
unchanged according to Ref.~\cite{prc_45_1098_1992}.
 Thus the modified tensor-even $\tilde{V}^{\rm TNE}$ and tensor-odd 
$\tilde{V}^{\rm TNO}$ interactions are described 
in relation to the parameter $\beta$ as 
\begin{subequations}
\begin{eqnarray}
\tilde{V}^{\rm TNE} & = & V^{\rm TNE} + 3(\beta-1)V^{\rm TNO} ,\\
\tilde{V}^{\rm TNO} & = & \beta V^{\rm TNO} ,
\end{eqnarray}
\label{eq_beta}
\end{subequations}
where $\tilde{V}^{\rm TNE}=V^{\rm TNE}$ and 
$\tilde{V}^{\rm TNO}=V^{\rm TNO}$ for $\beta=1$.
 The long-range part of the isovector tensor interaction is 
well known from the one-pion exchange model, however, the 
short-range part has not been determined as accurately.
 Thus we modified only the imaginary part of the short-range 
$V^{\rm TNE}$ and $V^{\rm TNO}$ interactions, which have a range of
0.25 fm, because the $D_{ii}(0^{\circ})$ value for the 
$J^{\pi}$ = $2^-$ state is sensitive to these components.

\begin{figure}
\begin{center}
\includegraphics[width=0.6\linewidth,clip]{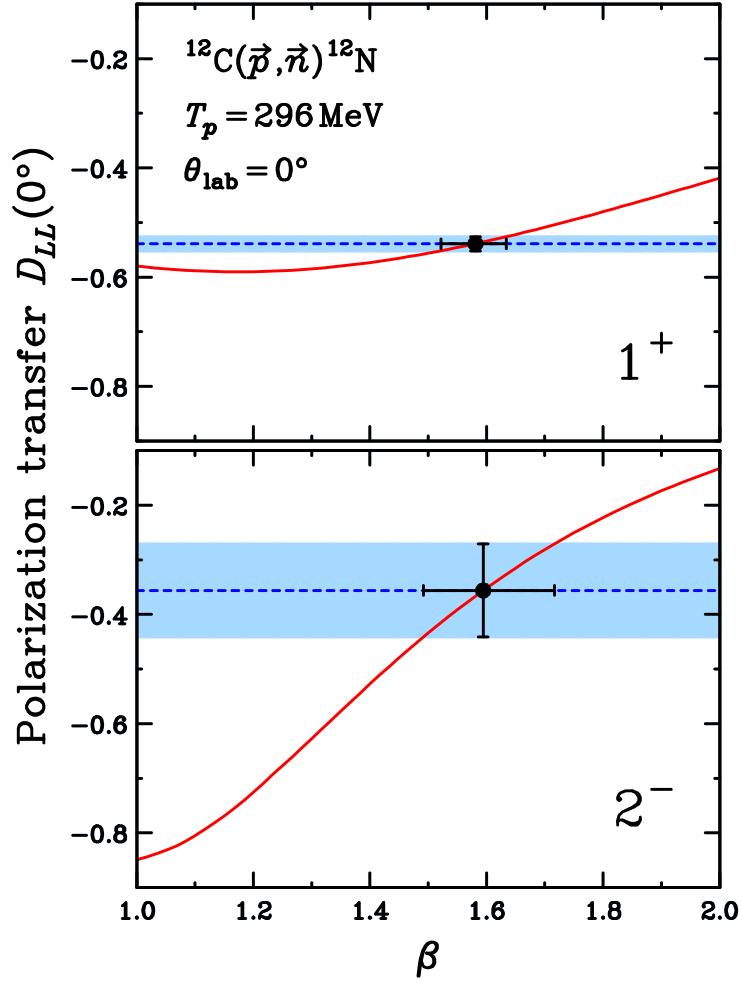}
\end{center}
\caption{(Color online) 
 The polarization transfer $D_{LL}(0^{\circ})$ for 
the 
${}^{12}{\rm C}(\vec{p},\vec{n}){}^{12}{\rm N}({\rm g.s.},1^+)$ 
(upper panel) and 
${}^{12}{\rm C}(\vec{p},\vec{n}){}^{12}{\rm N}(4.2\,{\rm MeV},2^-)$ 
(lower panel) 
reactions at $T_p$ = 296 MeV and $\theta_{\rm lab}$ = $0^{\circ}$ 
\cite{prc_80_024319_2009}.
 The data are shown by the filled circles and 
horizontal dashed lines, and 
the corresponding uncertainties shown by the vertical error bars 
and horizontal bands.
 The curves represent the results of the DWIA calculations 
as a function of $\beta$ as defined in Eq.~(\ref{eq_beta}).
 See the main text for details.}
\label{fig3}
\end{figure}

\begin{figure}
\begin{center}
\includegraphics[width=0.6\linewidth,clip]{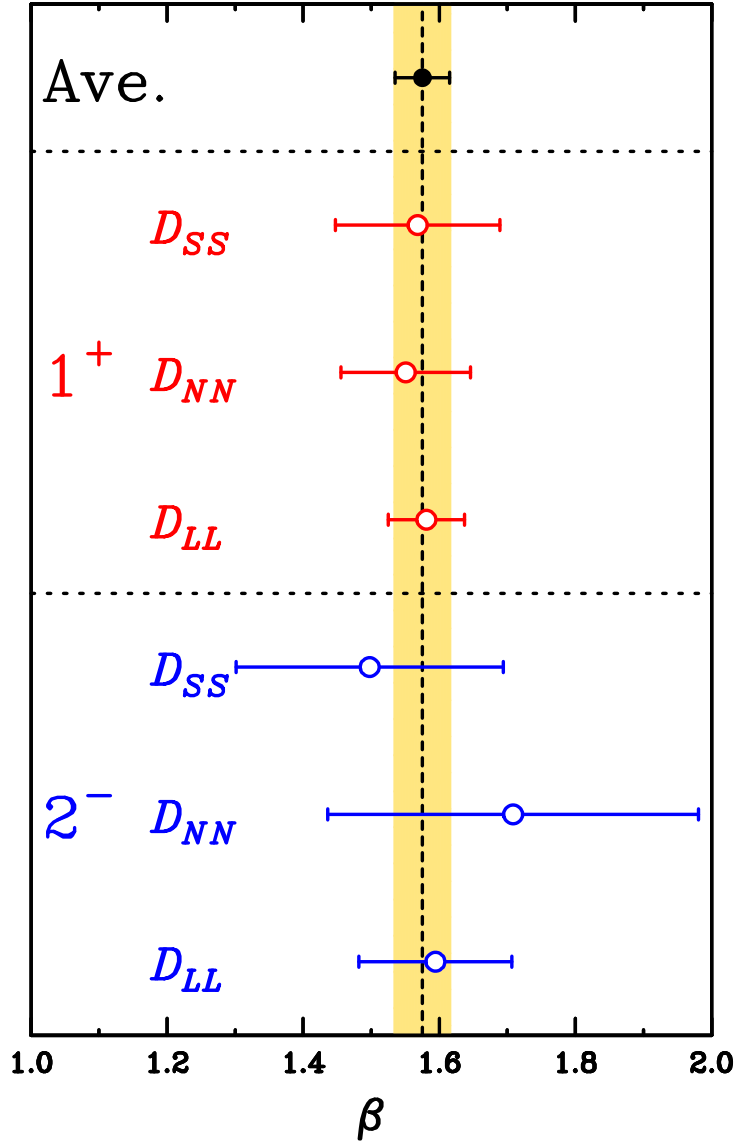}
\end{center}
\caption{(Color online) 
 Optimization of the $\beta$ values (see Eq.~(\ref{eq_beta})) 
to reproduce the polarization transfer observables 
$D_{ii}(0^{\circ})$ for the 
${}^{12}{\rm C}(\vec{p},\vec{n}){}^{12}{\rm N}({\rm g.s.},1^+)$ 
and 
${}^{12}{\rm C}(\vec{p},\vec{n}){}^{12}{\rm N}(4.2\,{\rm MeV},2^-)$ 
reactions at $T_p$ = 296 MeV and $\theta_{\rm lab}$ = $0^{\circ}$ 
\cite{prc_80_024319_2009}.
 The vertical line and band represent the averaged 
value and its uncertainty, respectively.}
\label{fig4}
\end{figure}

 The upper and lower panels of Fig.~\ref{fig3} represent
the $D_{LL}(0^{\circ})$ values for the $J^{\pi}$ = 
$1^+$ and $2^-$ states, 
respectively, as a function of $\beta$.
 The experimental data are shown by filled circles and 
horizontal dashed lines, and 
the corresponding uncertainties shown as vertical error bars 
and horizontal bands.
 The $D_{LL}(0^{\circ})$ values for both the $J^{\pi}$ = 
$1^+$ and $2^-$ states
are well reproduced using $\beta\simeq 1.6$ with 
uncertainties of $\delta\beta\simeq 0.1$. 
 The optimum $\beta$ values were also deduced for the other polarization 
transfers, $D_{SS}(0^{\circ})$ and $D_{NN}(0^{\circ})$, 
and the results are summarized in Fig.~\ref{fig4}.
 All the $D_{ii}(0^{\circ})$ data support the modification 
of the tensor component using $\beta\simeq 1.6$.
 In the following, therefore, 
we use the averaged value of $\beta = 1.58\pm0.04$
in the DWIA calculations.

\subsection{Comparison with theoretical calculations}

\begin{figure}
\begin{center}
\includegraphics[width=0.6\linewidth,clip]{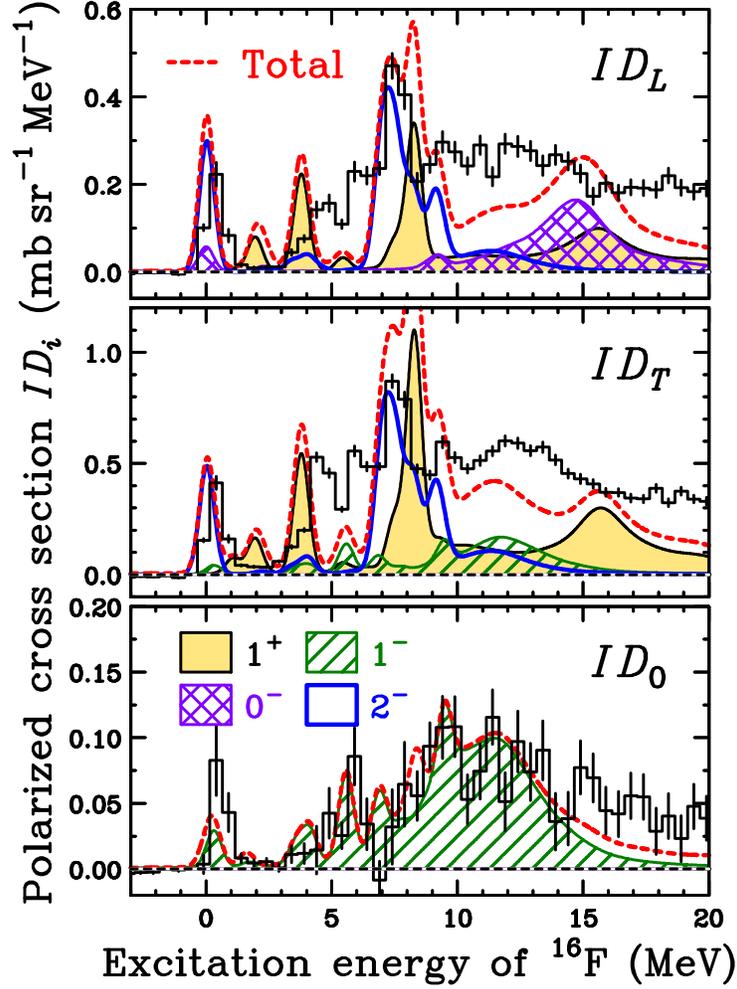}
\end{center}
\caption{(Color online) 
 The spin-longitudinal $ID_L$ (top panel), 
spin-transverse $ID_T$ (middle panel), and 
non-spin $ID_0$ (bottom panel) polarized 
cross sections for  
the ${}^{16}{\rm O}(\vec{p},\vec{n}){}^{16}{\rm F}$ reaction 
at $T_p$ = 296 MeV and $\theta_{\rm lab}$ = $0^{\circ}$.
 The shaded, cross-hatched, hatched, and unfilled regions 
represent the results of the DWIA calculations 
for the $J^{\pi}$ = $1^+$, $0^-$, $1^-$, and $2^-$ components, 
respectively.
 The dashed curves show the total $ID_i$ including contributions of up to 
$J^{\pi}$ = $4^-$.
 The intrinsic widths for the states at $E_x$ $\ge$ 9.5 MeV 
have been set to $\Gamma$ = 2 MeV.
 The DWIA results have been convoluted with a Gaussian function 
with an experimental energy resolution of 700 keV 
in FWHM.}
\label{fig5}
\end{figure}

 Figure~\ref{fig5} compares the experimental polarized 
cross sections $ID_i(0^{\circ})$ with the theoretical 
calculations.
 The intrinsic widths, $\Gamma$, have been neglected 
for the states at $E_x$ $<$ 9.5 MeV,
where narrow peaks and resonances are observed, 
whereas widths of $\Gamma$ = 2 MeV were used for the states 
at $E_x$ $\ge$ 9.5 MeV.
 The results of the calculations were convoluted with 
a Gaussian function with an experimental resolution of 700 keV 
in the full-width at half-maximum (FWHM).
 The shaded, cross-hatched, hatched, and unfilled regions 
correspond to the $J^{\pi}$ = 
$1^+$, $0^-$, $1^-$, and $2^-$ components, 
respectively.
 The total $ID_i(0^{\circ})$ spectrum, including components 
up to $J^{\pi}$ = $4^-$ 
are shown by the dashed curves, although the contributions 
from $L$ = 2--4 components are small.
 As expected in a simple PWIA \cite{prc_26_727_1982}, 
the spin-longitudinal cross section, $ID_L(0^{\circ})$, consists exclusively 
of the unnatural-parity $J^{\pi}$ = 
$1^+$, $0^-$, and $2^-$ transitions,
whereas the spin-transverse cross section, $ID_T(0^{\circ})$, 
consists of the unnatural-parity $J^{\pi}$ = $1^+$ and $2^-$ transitions 
as well as the natural-parity $J^{\pi}$ = $1^-$ transition.
 Note that the natural-parity $J^{\pi}$ = $1^-$ transition is 
predominant in the spin-scalar cross section, $ID_0(0^{\circ})$.

 The peaks at $E_x$ $\simeq$ 0 MeV in the $ID_L(0^{\circ})$ and 
$ID_T(0^{\circ})$ spectra
are reasonably reproduced as a combination of the 
dominant $J^{\pi}$ = $2^-$ contribution with 
weak $J^{\pi}$ = $0^-$ and $1^-$ contributions.
 The $J^{\pi}$ = $2^-$ state at $E_x$ $\simeq$ 7.5 MeV is also well 
reproduced in the present calculations.
 Furthermore, the SDR at $E_x$ $\simeq$ 9.5 MeV
observed in both the $ID_L(0^{\circ})$ and $ID_T(0^{\circ})$ spectra 
is reasonably reproduced by the calculations as 
the $J^{\pi}$ = $2^-$ state.
 Thus we conclude that 
the SDR at $E_x$ $\simeq$ 9.5 MeV is dominated by the 
$J^{\pi}$ = $2^-$ state.

 It is interesting that the $J^{\pi}$ = 
$2^-$ state at $E_x$ = 5.86 MeV
could not be reproduced.
 In the SM calculations, the 
$(0p_{1/2}^{-1}0d_{5/2})$, $(0p_{1/2}^{-1}0d_{3/2})$, 
$(0p_{3/2}^{-1}0d_{5/2})$, and $(0p_{3/2}^{-1}0d_{3/2})$ 
configurations are dominant for the 
experimental $J^{\pi}$ = 
$2^-$ states at $E_x$ = 0.42, 5.86, $\simeq$ 7.5, and
$\simeq$ 9.5 MeV, respectively.
 It should be noted that the $J^{\pi}$ = $2^-$ state 
at $E_x$ = 5.86 MeV 
is predicted to have a significant contribution 
from the $(0p_{3/2}^{-1}1s_{1/2})$ 
configuration.
 The interference between $(0p_{1/2}^{-1}0d_{3/2})$ and  
$(0p_{3/2}^{-1}1s_{1/2})$ reduces the transition strength, 
and thus the $ID_i(0^{\circ})$ becomes very small.
 This quenching is also exhibited in a standard SM calculation 
\cite{npa_637_547_1998}
using the PSDMKII interaction \cite{npa_255_315_1975}.
 Thus further detailed theoretical investigations are 
highly required to resolve the discrepancy 
for the $J^{\pi}$ = $2^-$ state at $E_x$ = 5.86 MeV.

\begin{figure}
\begin{center}
\includegraphics[width=0.6\linewidth,clip]{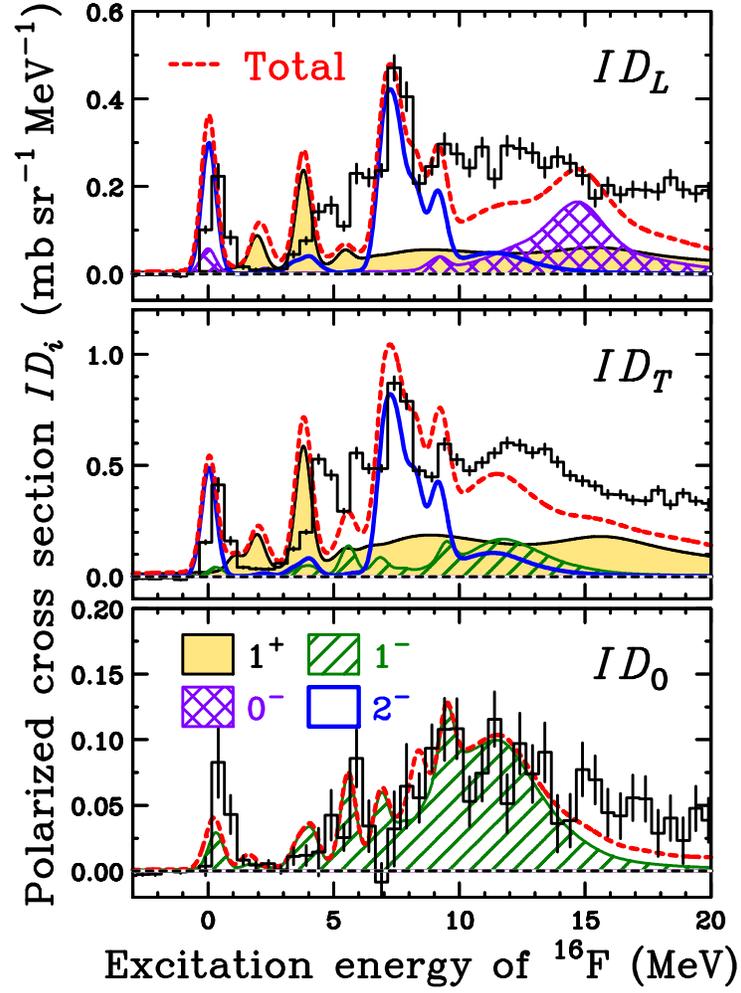}
\end{center}
\caption{(Color online) 
 Same as Fig.~\ref{fig5} but the intrinsic widths 
for the GT $1^+$ states at $E_x$ $\ge$ 7 MeV 
have been set to $\Gamma$ = 5 MeV.
 See text for details.}
\label{fig6}
\end{figure}

 For the $J^{\pi}$ = $1^+$ GT transitions, 
the calculations provide reasonable predictions of the magnitudes
of the peaks at $E_x$ = 3.76 and 4.65 MeV in 
both the $ID_L(0^{\circ})$ and $ID_T(0^{\circ})$ spectra, 
even though the excitation energies are 
significantly lower at $E_x$ $\simeq$ 1.9 and 3.8 MeV.
 The calculations also predict concentrations at 
$E_x$ $\simeq$ 8.3 and 15.7 MeV, which is inconsistent with the  
experimental data.
 A possible explanation for these features is that the strengths are
fragmented by enlarging the configuration space.  
Figure~\ref{fig6} shows the results of calculations 
with $\Gamma$ = 5 MeV for the GT states at $E_x$ $\ge$ 7 MeV, 
which provides a better description for 
both $ID_L(0^{\circ})$ and $ID_T(0^{\circ})$ values.
 The present data support the significant GT strengths 
predicted in the SM calculations.
 However, the calculations underestimate both 
$ID_L(0^{\circ})$ and $ID_T(0^{\circ})$  
at $E_x$ $\gtrsim$ 10 MeV.
 This underestimation might be resolved by considering the 
$4\hbar\omega$ configurations \cite{prl_65_1325_1990}.

 For the $J^{\pi}$ = $1^-$ SD transitions, 
two peaks which correspond to 
the states at $E_x$ = 0.19 and 5.24 MeV are clearly observed 
in the $ID_0(0^{\circ})$ spectrum.
 The observed state at $E_x$ = 5.24 MeV in $ID_0(0^{\circ})$ 
supports a tentative spin-parity assignment of the $J^{\pi}$ = $1^-$ state 
\cite{npa_564_1_1993}.
 The $ID_0(0^{\circ})$ peak for the state at $E_x$ = 5.24 MeV is well
reproduced by the theoretical calculations,
whereas that at $E_x$ = 0.19 MeV is 
underestimated, although the experimental uncertainty is 
large.
 The broad bumps in $ID_0(0^{\circ})$ and $ID_T(0^{\circ})$ 
at $E_x$ $\simeq$ 12 MeV 
are reasonably reproduced as the GDR and 
SDR, respectively.
 It should be noted that this broad bump is not observed in 
either the experimental or theoretical $ID_L(0^{\circ})$ spectrum, 
and thus it is natural 
to conclude that the bump is dominated by both $1^-$ SDR and GDR.

 Concerning the $J^{\pi}$ = $0^-$ SD transitions, 
the theoretical calculations 
predict both the well-established state at $E_x$ = 0 MeV and 
the missing SDR at $E_x$ $\simeq$ 15 MeV.
 The significant strength observed in $ID_L(0^{\circ})$ at around
$E_x$ = 15 MeV supports the existence of the $J^{\pi}$ = $0^-$ SDR 
in this region.
 However, the experimental data does not show a clear bump in this region, 
and thus the $J^{\pi}$ = $0^-$ strengths are likely to be more fragmented.
 Therefore, further detailed theoretical investigations are also 
required to determine the distribution of the $J^{\pi}$ = $0^-$ SDR.

\section{\label{sec:summary}SUMMARY AND CONCLUSION}

 The cross section and a complete set of polarization transfer 
observables were measured for the 
${}^{16}{\rm O}(\vec{p},\vec{n}){}^{16}{\rm F}$ reaction 
at $T_p$ = 296 MeV and $\theta_{\rm lab}$ = $0^{\circ}$.
 The experimental polarized cross sections $ID_i(0^{\circ})$ 
($i$ = 0, $L$, and $T$)
were compared with DWIA calculations, employing 
SM wave functions of up to $3\hbar\omega$ 
configurations.
 The GT and SD states at $E_x$ $\lesssim$ 8 MeV have been 
reasonably reproduced by the DWIA calculations, with the exception of  
the $J^{\pi}$ = $2^-$ state at $E_x$ = 5.86 MeV, in which the 
predicted contribution from the $(0p_{3/2}^{-1}1s_{1/2})$ 
configuration seems to be inappropriate.
 The SDR at $E_x$ $\simeq$ 9.5 MeV appears to have more 
$J^{\pi}$ = $2^-$ strength than $J^{\pi}$ = $1^-$ strength 
at $\theta_{\rm lab}$ = $0^{\circ}$, whereas 
the bump at $E_x$ $\simeq$ 12 MeV is reasonably explained 
as the sum of the $J^{\pi}$ = $1^-$ GDR and SDR.
 The data show a significant strength in $ID_L(0^{\circ})$ 
at $E_x$ $\simeq$ 15 MeV, 
which can be attributed to the $J^{\pi}$ = $0^-$ SDR predicted in
the SM calculations.
 These findings, and further studies applying polarization 
transfer measurements to other nuclei, will provide valuable 
insight for studies into nuclear structure, e.g., 
tensor correlations in nuclear spin excitations.

\begin{acknowledgments}
 We are grateful to Professor T. Kawabata for his helpful 
correspondence.
 We also acknowledge the dedicated efforts 
of the RCNP cyclotron crew for providing a high-quality 
polarized proton beam.
 The experiment was performed at RCNP under Program Number 
E317.
 This research was supported in part by the Ministry of 
Education, Culture, Sports, Science, and Technology of Japan.
\end{acknowledgments}

\clearpage

\clearpage

\clearpage

\clearpage

\clearpage

\clearpage

\clearpage

\bibliography{e317}

\end{document}